\documentclass[conference,9pt]{IEEEtran}
\AtBeginDocument{%
  \providecommand\BibTeX{{%
    \normalfont B\kern-0.5em{\scshape i\kern-0.25em b}\kern-0.8em\TeX}}}
\usepackage{amsmath,amsfonts}
\usepackage{algorithm}
\usepackage{algorithmicx}
\usepackage{algpseudocode}

\usepackage{graphicx}
\usepackage{subcaption}
\usepackage{tikz}
\newcommand*\circled[1]{\tikz[baseline=(char.base)]{
            \node[shape=circle,draw,inner sep=0.6pt] (char) {#1};}}
\usepackage[export]{adjustbox}
\usepackage{pdfpages}
\usepackage{textcomp}
\usepackage{xcolor}
\def\BibTeX{{\rm B\kern-.05em{\sc i\kern-.025em b}\kern-.08em
    T\kern-.1667em\lower.7ex\hbox{E}\kern-.125emX}}

\usepackage{ulem}
\usepackage{braket}
\usepackage{mathtools}
\DeclarePairedDelimiter{\ceil}{\lceil}{\rceil}
\usepackage{enumitem}
\newcommand\blfootnote[1]{%
  \begingroup
  \renewcommand\thefootnote{}\footnote{#1}%
  \addtocounter{footnote}{-1}%
  \endgroup
}
\renewcommand{\footnoterule}{%
\kern-3pt 
\hrule height .3pt width 1.4in \kern 2.6pt
}

\begin{document}

\title{A Scalable and Robust Compilation Framework for \\Emitter-Photonic Graph State}
\author{\IEEEauthorblockN{
Xiangyu Ren\IEEEauthorrefmark{2}\IEEEauthorrefmark{1},
Yuexun Huang\IEEEauthorrefmark{3}, 
Zhiding Liang\IEEEauthorrefmark{4}\IEEEauthorrefmark{1},
Antonio Barbalace\IEEEauthorrefmark{2}}
\IEEEauthorblockA{
\IEEEauthorrefmark{2}
University of Edinburgh, Edinburgh, Scotland, UK\\
\IEEEauthorrefmark{3}
University of Chicago, Chicago, IL, USA\\
\IEEEauthorrefmark{4}
Rensselaer Polytechnic Institute, Troy, NY, USA
}}
\maketitle

\begin{abstract}
Quantum graph states are critical resources for various quantum algorithms, and also determine essential interconnections in distributed quantum computing.
There are two schemes for generating graph states | \textit{probabilistic scheme} and \textit{deterministic scheme}. While the all-photonic \textit{probabilistic scheme} has garnered significant attention, the emitter-photonic \textit{deterministic scheme} has been proved to be more scalable and feasible across several hardware platforms.

This paper studies the GraphState-to-Circuit compilation problem in the context of the \textit{deterministic scheme}. 
Previous research has primarily focused on optimizing individual circuit parameters, often neglecting the characteristics of quantum hardware, which results in impractical implementations. Additionally, existing algorithms lack scalability for larger graph sizes.
To bridge these gaps, we propose a novel compilation framework that partitions the target graph state into subgraphs, compiles them individually, and subsequently combines and schedules the circuits to maximize emitter resource utilization. Furthermore, we incorporate \textit{local complementation} to transform graph states and minimize entanglement overhead. Evaluation of our framework on various graph types demonstrates significant reductions in CNOT gates and circuit duration, up to 52\% and 56\%. Moreover, it enhances the suppression of photon loss, achieving improvements of up to $\times$1.9. \blfootnote{\IEEEauthorrefmark{1} Corresponding to: Zhiding Liang (liangz9@rpi.edu) and Xiangyu Ren (xiangyu.ren@ed.ac.uk).}
\end{abstract}

\section{Introduction}
Quantum information processing has the potential to revolutionize computation, communication, cryptography, sensing, and imaging. Within quantum information, entanglement plays a vital role as it serves as the foundation for interconnections between qubits. 
\textit{Graph state} -- a highly entangled quantum state, is the critical resource for quantum error correction (QEC)~\cite{schlingemann2001quantum}, measurement-based quantum computing (MBQC)~\cite{raussendorf2001one}, and various other applications.
Moreover, it enables distributed quantum computing~\cite{liu2024ecdqc,ferrari2023simulation} and quantum networking~\cite{donne2024design}, acting as a paradigm for non-local qubit interactions. For instance, the \textit{Bell state} is a special case of a graph state.

Before graph states can be utilized for various applications, they must first be generated on quantum computers or quantum network nodes. Graph state generation has predominantly focused on the \textit{probabilistic scheme}~\cite{browne2005resource}, which relies on a sequence of \textit{fusion gates} and photon interference. This all-photonic approach is widely studied for its relevance in optical quantum computing. Several previous compilation studies~\cite{oneperc,li2023minimizing,MoFCM} for photonic MBQC are rooted in the \textit{probabilistic scheme}. However, such approach suffers from an exponential demand for resources, severely limiting the scalability of the target graph states. Additionally, the randomness and uncertainty associated with fusion gates significantly increase the complexity of the compilation algorithms.

An alternative hardware solution is the \textit{deterministic scheme}, where \textit{photon qubits} are directly generated by emitters~\cite{russo2018photonic}. In this scheme, a set of interacting \textit{emitter qubits} is employed to induce entanglement among photon qubits as they are emitted. Owing to better scalability and advancements in engineering, the \textit{deterministic scheme} is attracting increasing attention. Emitter-based graph states have been demonstrated across various hardware platforms, including silicon quantum dots~\cite{arakawa2020progress,cogan2023deterministic}, nitrogen vacancy (NV) color centers~\cite{nemoto2014photonic,choi2019percolation}, and Rydberg atoms~\cite{yang2022sequential}, achieving scales of up to 14 qubits~\cite{thomas2022efficient}.

\begin{figure}[!t]
% \vspace{3mm}
    \centering
    \includegraphics[page=1,width=.4\textwidth]{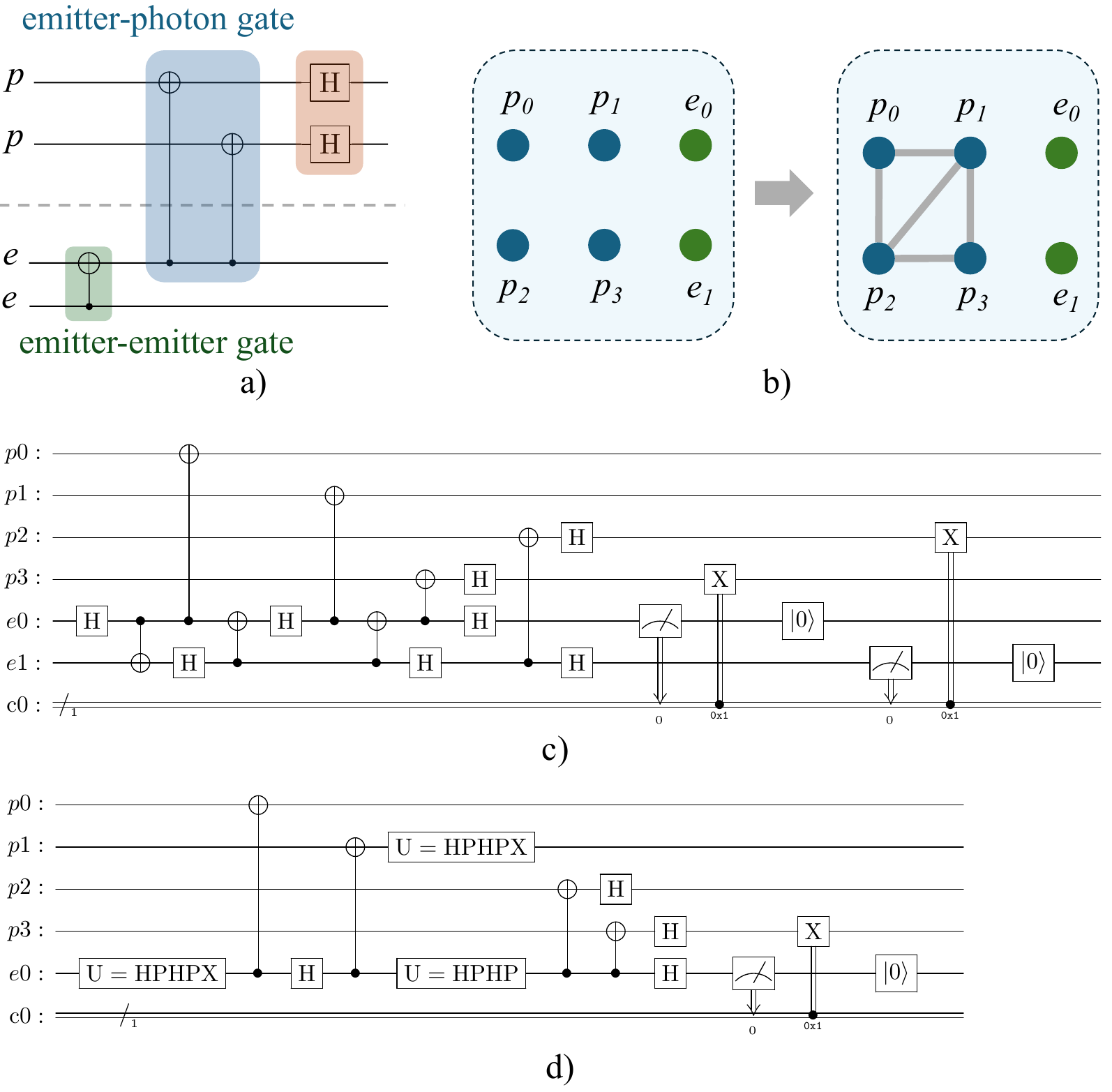}
    \caption{The background on GraphState-to-Circuit Compilation. a) The constraints of a \textit{deterministic} graph state generation, with $p$ as photon and $e$ as emitter. For deterministic purpose, only emitter-emitter CNOT in green and emitter-photon CNOTs (emissions) in blue are allowed. b) Generation of a graph state on photon qubits $p_0$ - $p_3$. c) The circuit for generating the graph state in b), as a result of compilation. d) An optimized circuit for the same graph state, with fewer emitters utilization and fewer CNOTs engaged compared to c).} 
    \label{fig:circmodel}
\vspace{-5mm}
\end{figure}

The generation process of an emitter-based graph state could be formalized by a quantum circuit model.
Constructing the quantum circuit for the generation of a specific graph state can be referred to as the \textbf{GraphState-to-Circuit compilation} | which is the problem we focus on in this paper.
Such a compilation task has distinct constraints compared to previous quantum compilers designed for superconducting~\cite{li2019tackling,tannu2019ensemble,tannu2019not,molavi2022qubit,shi2019,ren2024mlqc}, trapped-ion\cite{tseng2024satisfiability}, all-photonic~\cite{oneperc}, Bosonic quantum~\cite{zhou2024bosehedral}, or neutral atom~\cite{wang2024atomique} platforms.
Shown in Figure~\ref{fig:circmodel}.a, the circuit model for graph state generation~\cite{li2022photonic,ghanbari2024optimization} contains two types of qubits: \textit{emitter qubit} and \textit{photon qubit}.
Entanglement is created via \uline{emitter-emitter interactions} (e.g., emitter-emitter CNOTs), and the entanglement connections are transduced to photons through the \uline{photon emission} process.
However, for the purpose of being \textit{deterministic}, \uline{photon-photon interactions are not allowed}, which significantly differentiates this approach from the all-photonic \textit{probabilistic scheme}.
Moreover, the gates for different types of qubits are implemented based on completely different physical operations, leading to \uline{significantly varying gate durations}.
These particular constraints necessitate a specific compiler.

A series of research efforts attempt to address the compilation problem for emitter-based graph state generation.
Li et al.~\cite{li2022photonic} propose a method to minimize the emitter resources required for graph state generation.
However, their approach does not consider circuit depth and suffers from photon loss issues due to the prolonged hardware operations of emitter CNOTs.
Kaur et al.~\cite{kaur2024resource} explore loss-aware generation for repeater graph states (RGS), but their method does not fully exploit resource utilization for arbitrary graph states.
Ghanbari et al.~\cite{ghanbari2024optimization} leverage \textit{local complementation} to optimize graph states and reduce generation overhead.
Nevertheless, finding the optimal \textit{local complementation} is \#P-complete~\cite{ji2024distributing}, which limits the scalability of this optimization.
Furthermore, none of these works provide a feasible compilation scheme for large-scale graphs.
Hindered by high complexity and an exponential solution space, optimal solutions for random graph states can only be achieved at the scale of a few qubits.

To address the above limitations, we propose a novel compilation framework for the graph state generation task, adopting a divide-and-conquer strategy to tackle the scalability problem.
First, we partition the graph state into subgraphs, enabling optimal compilation for each subgraph with low overhead, while integrating \textit{local complementation} for optimization.
Next, we compile the subgraphs based on a comprehensive objective function that accounts for the characteristics of actual quantum computing hardware.
Finally, during subgraph recombination, we design a circuit scheduling scheme to exploit emitter reuse among subgraphs, thereby maximizing qubit resource utilization and reducing photon loss accumulation during generation.
Our framework is evaluated on various types of graph states and compared with the best-known method~\cite{lin2024graphiq}.
It outperforms the baseline, achieving on average a 30\% (up to 52\%) reduction in emitter-emitter CNOTs, a 38\% (up to 56\%) reduction in circuit duration, and a suppression of photon loss by $\times$1.4 (up to $\times$1.9).

Our key contributions are as follows:
\begin{enumerate}
    \item We are the first to study emitter-based graph state generation from a quantum compiler perspective. We formulate the compilation constraints, perform a comprehensive analysis of previous methods, and identify the objectives to consider based on realistic hardware implementations.
    \item We propose a scalable compilation method by partitioning the target graph state into subgraphs and compiling these subgraphs. This method incorporates \textit{local complementation} to reduce the critical overhead -- emitter-emitter CNOT.  
    \item We design a scheduling scheme that maximizes the utilization of emitter qubit resources. It minimizes circuit duration, thereby reducing the accumulation of photon loss.  
\end{enumerate}

\section{Background}
\subsection{Quantum Graph State}
A quantum graph state $\ket{G}$ is a specific type of multipartite entanglement state that is intrinsically associated with a graph $G=(V,E)$, where $V$ is the set of vertices and $E$ is the set of edges connecting these vertices. In graph $G$, a vertex $v$ represents a physical qubit, whereas an edge $e$ indicates the entanglement between the two qubits.
The formal definition of a graph state $\ket{G}$ is given by:
\begin{equation}
    \ket{G} = \prod_{(i,j)\in E}{CZ_{(i,j)}\ket{+}^{\otimes \lvert V \rvert}}
\end{equation}
Here, $\ket{+}^{\otimes \lvert V \rvert}$ denotes the tensor product of $\lvert V \rvert$ copies of $\ket{+}$ state at all the vertices of $G$, and $CZ_{(i,j)}$ is the controlled-Z gate operation applied to the pair of qubits corresponding to the edge $(i,j) \in E$. 

Typically, the generation of a graph state is achieved through two steps: 
1) Prepare a set of qubits in initial states, corresponding to all the vertices in $V$; 
2) For each pair of qubits where the entanglement is desired, apply a CNOT/CZ interaction. 
Note that for a graph state, all CNOT/CZ gates are commutable, allowing them to be arranged in arbitrary sequences or applied in parallel, providing abundant flexibility for circuit scheduling.
Here we provide an example of graph state, shown on the right side of Figure~\ref{fig:circmodel}.b), representing the entanglement among photon qubits $p_0,p_1,p_2,p_3$.

\subsection{Emitter-Based Graph State Generation}
\textbf{Problem Formulation.}
Here we provide a description of the circuit model for emitter-based graph state generation, including all elementary operations required for the process, as illustrated in Figure~\ref{fig:circmodel}.a-c, and an example of optimization in Figure~\ref{fig:circmodel}.d:
\begin{enumerate}
    \item Two types of qubits are involved: \textit{emitter qubit} and \textit{photon qubit}. Emitter qubits are initialized to state $\ket{0}$, while photon qubits do not exist physically before emission and are represented as $\ket{0}$.
    
    \item Entanglement is induced by emitter-emitter CNOT, and the entanglement connection are transduced to emitted photon via emission process.

    \item The emission of a photon is modeled by an emitter-photon CNOT between the emitter and emitted photon, being always the first gate applied on each photon qubit.

    \item For \textit{deterministic} purpose, the photon-photon interactions are not allowed. Only single-qubit gates and measurement can be applied on emitted photon qubits.

    \item The final result should be a graph state entangled across $n$ photons ($n$ = size of the graph), while all emitters return to $\ket{0}$ state upon completion.
\end{enumerate}

\textbf{Hardware Characteristic.} The emission and entanglement operations are implemented differently across hardware platforms. Here we use the quantum dots platform~\cite{russo2018photonic,gimeno2019deterministic} as an example.
The emitter-emitter CNOT is realized through electron exchange coupling between quantum dots, with a gate time of 1-10 ns and a fidelity $\geq$0.99. 
In contrast, the emitter-photon CNOT (emission) has a significantly shorter duration of $\sim$0.1 ns.
Given this distinction in gate time, circuit scheduling for emitter-based graph state generation becomes a non-trivial problem.

\begin{figure}[b]
\vspace{-5mm}
    \centering
    \includegraphics[page=1,width=.45\textwidth]{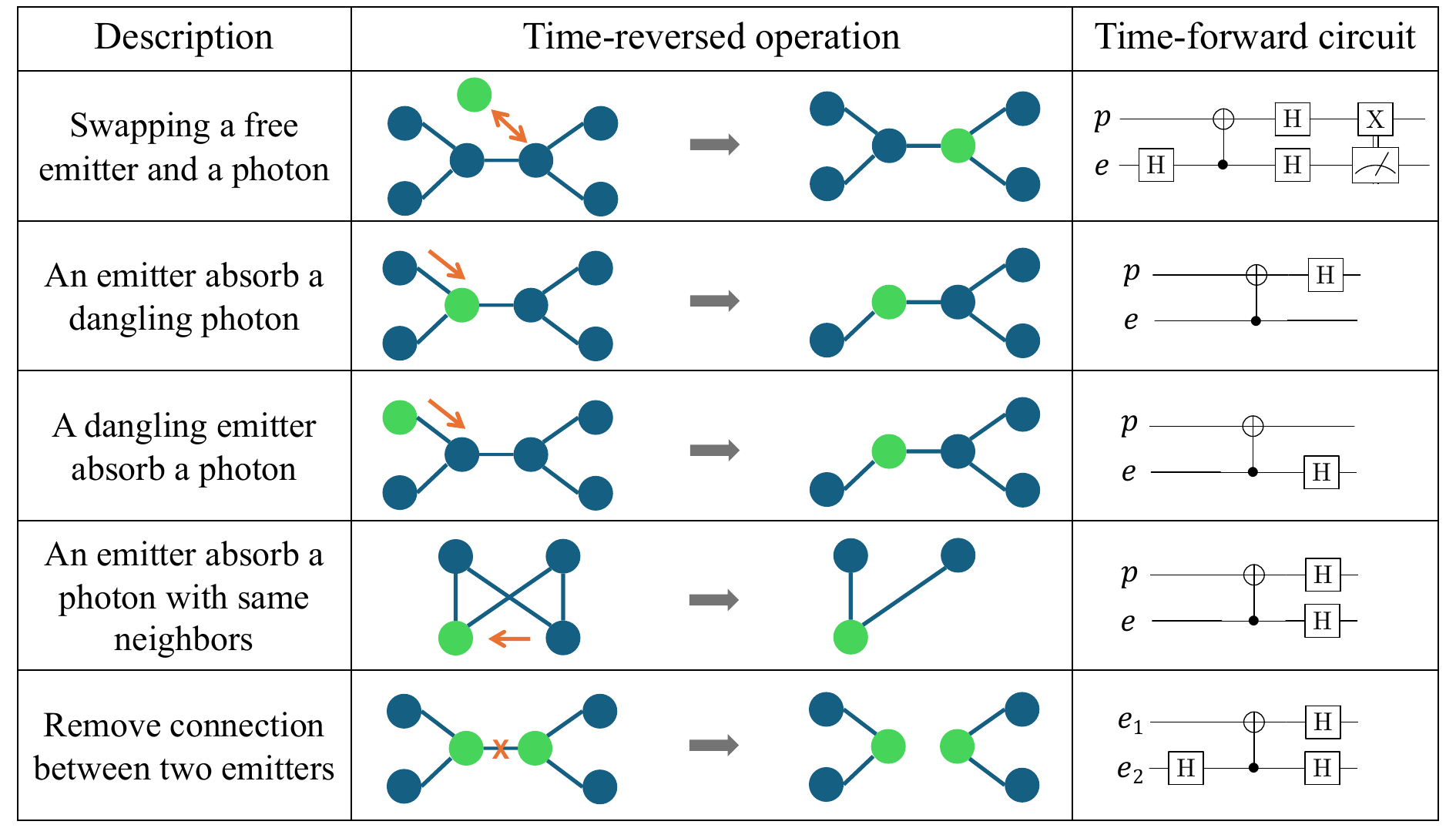}
    \caption{Operations of the time reversed graph reduction model.} 
    \label{fig:reverseop}
% \vspace{-3mm}
\end{figure}

\begin{figure}[t]
% \vspace{3mm}
    \centering
    \includegraphics[page=1,width=.5\textwidth]{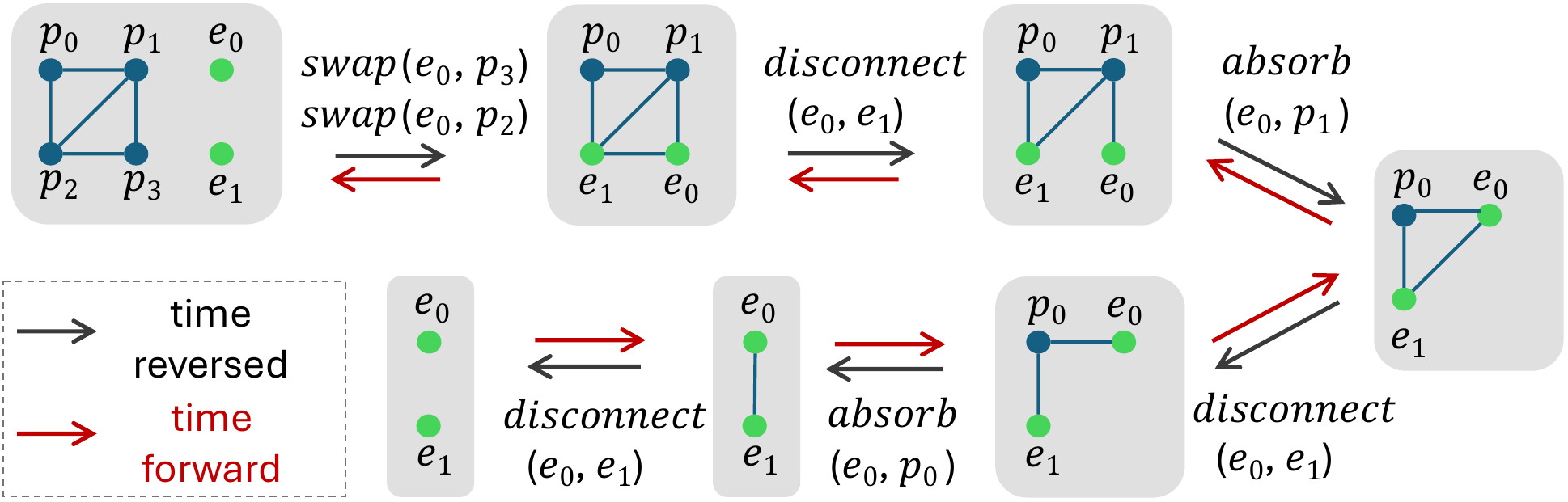}
    \caption{An example of the time-reversed reduction sequence (black arrows), corresponding to the circuit in Figure~\ref{fig:circmodel}.c (red arrows).} 
    \label{fig:reverseexample}
\vspace{-5mm}
\end{figure}

\subsection{Time-Reversed Model for Graph State Generation}
Our compilation algorithm is built upon a \textbf{time-reversed model} introduced in previous work~\cite{li2022photonic}. 
In this time-reversed context, photon emission is replaced with a hypothetical "photon absorption", while emitter-emitter interaction correspond to "disentangling" between emitters.
Using these reversed operations, we can derive a virtual sequence that reduces a target graph state to initial state. Hence, reversing this sequence yields the required circuit to generate target state from initial state. We specify the following reversed operations~\cite{kaur2024resource}, which are also illustrated in Figure~\ref{fig:reverseop}:

\paragraph{Swapping with free emitter} An arbitrary photon in the graph state can be replaced by a free emitter in $\ket{0}$ state.

\paragraph{Photon absorption} If an emitter is entangled with a photon in the graph, it can absorb the photon under certain conditions. The specific conditions for photon absorption are illustrated in Figure~\ref{fig:reverseop}.

\paragraph{Emitter disentangle} Remove the entanglement connection (represented as an edge in the graph) between two emitters.

Based on these operations, our compiler determines the optimal time-reversed reduction sequence of target graph state (e.g. Figure~\ref{fig:reverseexample}), then reverses this sequence to generate the desired circuit. We recommend Kaur et al.~\cite{kaur2024resource} for a more comprehensive background.

\subsection{Optimizing Graph State via Local Complementation}
By applying local Clifford (LC) gates to a graph state, it can be transformed into a set of alternative graph states that are LC-equivalent with original graph state. 
Hence, generation of the target graph state can be redefined as generating an LC-equivalent graph state. If the LC-equivalent graph has fewer edges, this transformation can reduce the entanglement overhead, with the only cost being the addition of single-qubit gates to the circuit.

\begin{figure}[h]
\vspace{-3mm}
    \centering
    \includegraphics[page=1,width=.35\textwidth]{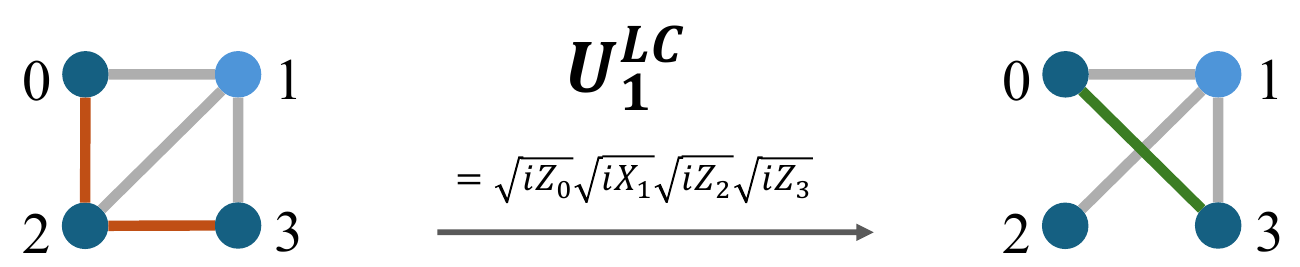}
    \caption{Leveraging local complementation to optimize the graph state. In this example, a Clifford unitary is applied on photon 1 ($U_1^{LC}$), leading to a local complementation of the corresponding vertex 1, acting on neighborhood vertices 0, 2, 3.} 
    \label{fig:localc}
\vspace{-3mm}
\end{figure}

Finding LC-equivalent graph state can be formalized as applying \textit{local complementation} to the graph~\cite{ghanbari2024optimization,ji2024distributing}. Specifically, local complementation of a vertex $v$ in the graph modifies the edges within the neighborhood on $v$. In the neighborhood of $v$, edges that are present will be removed, and edges that are missing will be added (Figure~\ref{fig:localc}). A proper local complementation of target graph state can significantly reduce the entanglement overhead. However, determining the optimal local complementation has been proven to be \#P-complete~\cite{dahlberg2020counting}, making it infeasible to solve in polynomial time.

\section{Motivation}
\subsection{Challenges and Solution}
Given the constraints and optimization opportunities, GraphState-to-Circuit compilation is a complex and non-trivial task. In this section, we discuss the associated challenges and propose our solution.

\begin{center}
    \fcolorbox{black}{gray!10}{\parbox{.95\linewidth}
    {Challenge 1: There is an exponential overhead in searching the optimal circuit of graph state generation. Additionally, finding the best local complementation for target graph state is a \#P-complete problem. As a result, the compilation is \textbf{not scalable}.}}
\end{center}

As we mentioned in Section II.A, the commutation of CNOT/CZ in graph state generation allows photon emission in an arbitrary sequence. As a result, the compilation overhead grows exponentially when exhaustively searching all  possible circuits. 
For the state-of-the-art compiler GraphiQ~\cite{lin2024graphiq}, the runtime exceeds $10^3$ seconds when solving a linear cluster state with more than 10 qubits. Furthermore,  when realistic optimization targets such as circuit duration and loss rate are considered instead of simply minimizing the \#CNOT, the runtime becomes even worse.

\textit{Solution}: We partition the target graph state into several subgraphs, each with a feasible size for finding the optimal generation circuit. These subgraphs are treated as \textit{leaves} and are compiled separately with designated optimization. Meanwhile, those inter-subgraph entanglements are treated as \textit{stems}, compiled into an independent circuit and integrated into the final result.

\begin{center}
    \fcolorbox{black}{gray!10}{\parbox{.95\linewidth}
    {Challenge 2: In hardware-aware scenarios, there are \textbf{various factors affecting the quality} of compiled circuits. }}
\end{center}

Based on the hardware characteristic of quantum dots emitter in Section II.B, several optimization targets must be considered: 1) A limited number of emitters, leading to prolonged circuit execution; 2) Accumulation of photon loss over time, which may render photons unusable; 3) Imperfect emitter-emitter CNOTs, affecting fidelity of the target state. Previous research has only considered one of these factors~\cite{li2022photonic}, resulting in impractical outcomes.

\textit{Solution}: We construct a cost function that combines multiple optimization objectives. Specifically, we extract the gates duration based on hardware settings, and use them to calculate the photon loss duration of generation circuit as the cost function.

\begin{figure}[t]
% \vspace{3mm}
    \centering
    \includegraphics[page=1,width=.4\textwidth]{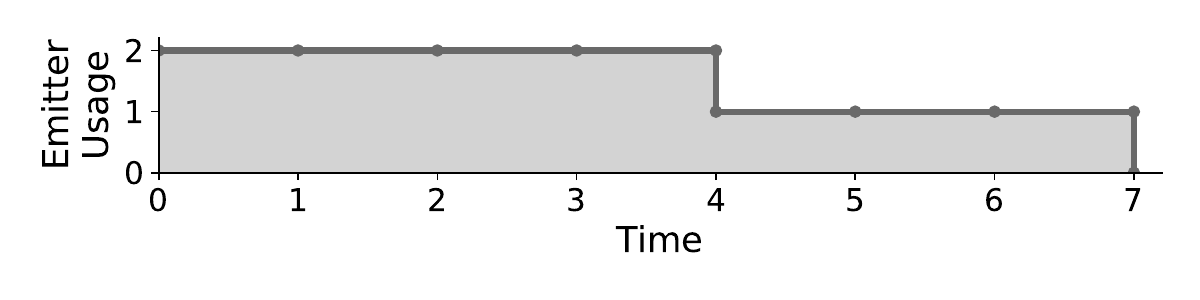}
    \vspace{-2mm}
    \caption{The emitter usage number over time (timescale simplified), for a graph state generation circuit.} 
    \label{fig:emitterusage}
\vspace{-3mm}
\end{figure}

\begin{center}
    \fcolorbox{black}{gray!10}{\parbox{.95\linewidth}
    {Challenge 3: When an emitter is entangled in the graph state, it may remain idle due to the lack of available operations. This results in \textbf{insufficient utilization of emitters}, thereby increasing the circuit duration.}}
\end{center}

Figure~\ref{fig:emitterusage} provides an example of the emitter usage curve of a circuit, illustrating the number of utilized emitters during the generation process. It can be observed that, for certain parts of the circuit, the utilization of emitters does not reach the maximum capacity. In other words, the emitter resources are not fully utilized. This under-utilization becomes more pronounced as the scale of the target graph state increases, resulting in a generation circuit with significantly longer duration.

\textit{Solution}: We schedule the subgraphs circuits by arranging them on the timeline. Based on the emitter utilization curve of each subgraph, we can share emitter qubits among different subgraphs, ensuring maximum utilization of emitters at every time slot. Briefly, we exploit the circuit-level parallelism and emitter reuse.

\begin{figure*}
\vspace{-3mm}
    \centering
    \includegraphics[page=1,width=0.9\textwidth]{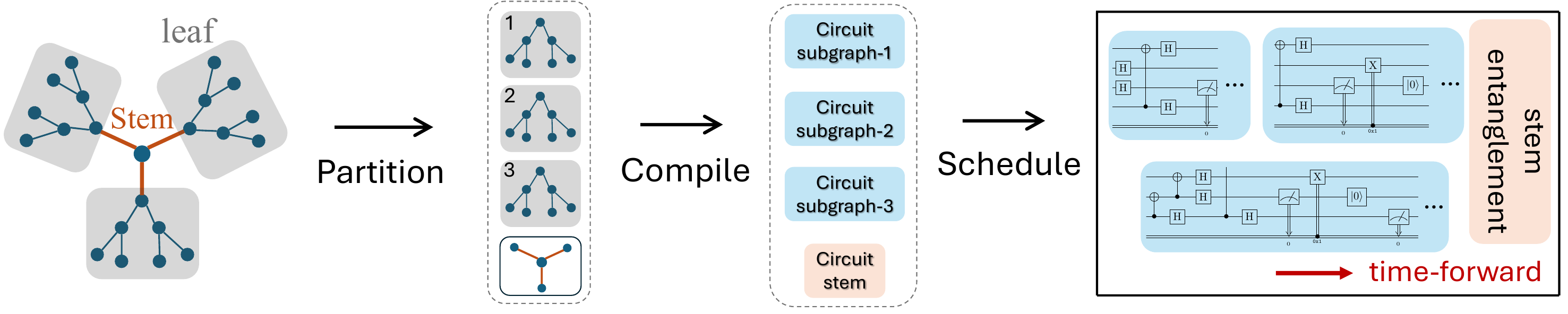}
    \caption{The overview of our framework.} 
    \label{fig:overview}
\vspace{-3mm}
\end{figure*}

\subsection{Insight: Divide-and-Conquer}
The framework we propose adopts a \textit{divide-and-conquer} strategy, inspired by the natural \textit{stem-leaf} structure of a tree graph, but also applies to arbitrary graph structure.  
A straightforward illustration of this \textit{stem-leaf} structure can be find in the left part of Figure~\ref{fig:overview}.
By partitioning the target graph state into several subgraphs~\cite{ren2024hardware}, we treat the entanglements within each \textit{leaf} as sub-problems, and finally recombine them through the compilation of inter-subgraph entanglements as the \textit{stem}. The overview of our framework is shown in Figure~\ref{fig:overview}.

Our insight within this strategy is that finding the global optimal solution of graph state generation circuit is highly complex. However, optimizing each sub-problem individually can reduce overhead and achieve a local-minimum. 
Another challenge is determining an appropriate partitioning scheme for any graph state, which will be discussed in Section~IV.A.
Additionally, our circuit scheduling approach reduces redundancy by treating each subgraph as a coarse-grained basic unit.

% \vspace{-3mm}
\section{Framework Design \& Implementation}
\subsection{Graph State Partitioning}
We present the design of the first part in our compilation framework, where the target graph state is partitioned into subgraphs. To find an appropriate partitioning scheme, the following considerations should be addressed:

\textbf{Reduce interconnections among subgraphs.} 
The target graph state is partitioned into subgraphs, and each subgraph is compiled with lower runtime overhead.
After obtaining the optimal generation circuit for each subgraph, they need to be connected via inter-subgraph entanglements (edges).
In the resulting circuit that combines all subgraphs, these inter-subgraph entanglements will be compiled into emitter-emitter CNOTs, which are expensive operations on the hardware.
Thus, the objective of graph state partitioning is to minimize the number of interconnection edges among different subgraphs.

\begin{figure}[b]
\vspace{-5mm}
    \centering
    \includegraphics[page=1,width=.42\textwidth]{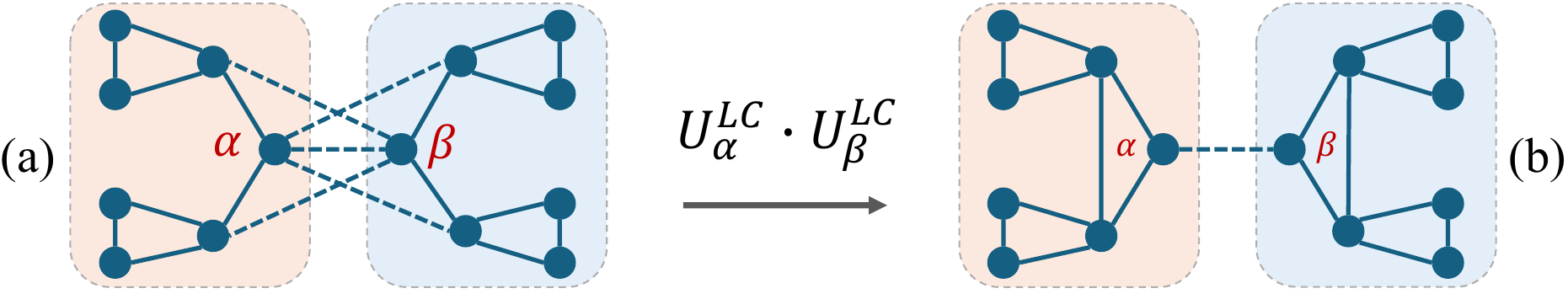}
    \caption{LC optimization for graph state partitioning.} 
    \label{fig:lcpartition}
% \vspace{-3mm}
\end{figure}

\textbf{Optimize graph partitioning with local complementation.}
Taking the graph state in Figure~\ref{fig:lcpartition} as an example,
normally at least 5 interconnecting edges must be cut to divide it into two subgraphs, shown in (a). 
However, by applying local complementation on vertices $\alpha$ and $\beta$, only 1 interconnecting edge remains for the partitioning, shown in (b). 
Thus, finding the appropriate local complementation is crucial not only for reducing the overall number of edges, but also for minimizing the partitioning cost.
Due to the exponential search space for local complementations, we limit the number of LC operations to $l$, resulting in a depth-limited search.
This restriction not only simplifies the search process, but also limits the additional circuit depth introduced by the local complementation operations.

\textbf{Graph state partition searcher with depth-limited LC.} We leverage a mix integer programming (MIP) solver to simultaneously find the best local complementation and graph partitioning solution. Our MIP model is described as follows:

\paragraph{Constant Parameters}
Given the target graph state $G=(V,E)$, the vertices are represented as $v \in V$ and the edges as $e \in E$. 
Assuming the number of local complementation (LC operations) is limited to $l$, we define a sequence $1,2,\ldots,l$ to mark the status of the graph state at each step. The subgraphs partitioned from $G$ are represented as the set $\{g\}$, with size limitation $|g|\leq g_{max}$. Additionally, the total number of subgraphs is given by $\ceil{|G|/g_{max}}$.

\paragraph{Variables}
First, we define our edge status in each step $t$ by $e_{v_1,v_2,t}$ ($t\in{1,\dots,l}$), and vertices in each subcircuit by $y_{v,g}$:
\vspace{-1mm}
$$ 
e_{v_1,v_2,t}=\left\{
\begin{aligned}
1 & \text{, if vertex } v_1, v_2 \text{ are adjacent at step } t \\
0 & \text{, otherwise} \\
\end{aligned}
\right.
$$
\vspace{-1mm}
$$ 
y_{v,g}=\left\{
\begin{aligned}
1 & \text{, if vertex } v \text{ is partitioned in subgraph } g \\
0 & \text{, otherwise} \\
\end{aligned}
\right.
$$
Then, we set the local complementation in each step as $x_{v,t}$.
\vspace{-1mm}
$$ 
x_{v,t}=\left\{
\begin{aligned}
1 & \text{, if local complementation applied on } v \text{ at step } t \\
0 & \text{, otherwise} \\
\end{aligned}
\right.
$$

\paragraph{Constraints}
For local complementation, we only apply it on single vertex at each step (or skip this step), thus we have:
\vspace{-1mm}
\begin{equation}
    \sum_{v \in V}{x_{v,t} \leq 1}, ~~~\forall t \in \{1,...,l\}
\end{equation}
Most importantly, the local complementation at step $t$ will add or remove the edges, resulting in a change at step $t+1$. Suppose the local complementation is applied to vertex $v_0$ ($x_{v_0,t}=1$), and $v_1, v_2$ being two of its adjacent vertices ($e_{v_0,v_1,t} = e_{v_0,v_2,t}=1$). Then, there should be a change to the edge $e_{v_1,v_2}$ from step $t$ to $t+1$:
\vspace{-1mm}
\begin{equation}
    x_{v_0,t} * e_{v_0,v_1,t} * e_{v_0,v_2,t}
    \leq |e_{v_1,v_2,t} - e_{v_1,v_2,t+1}|
\end{equation}
\vspace{-1mm}
In addition, the size of each subgraph is restricted by:
\begin{equation}
    \sum_{v \in V}{y_{v,g} \leq g_{max}}, ~~~\forall g \in \{g\}
\end{equation}
\vspace{-2mm}
\paragraph{Objective Function}
Since we hope to minimize the interconnecting edges among subgraphs, our objective function is the number of such edges, represented by $K$. We use $\sum_{\forall g \in \{g\}}{y_{v_1,g} * y_{v_2,g}}$ to determine if $v_1,v_2$ are in the same subgraph, then we have:
\begin{equation}
K=
\sum_{\forall v_1,v_2 \in V}{
    e_{v_1,v_2,l} * 
    (1-\sum_{\forall g \in \{g\}}{y_{v_1,g} * y_{v_2,g}})}
\end{equation}

Overall, the MIP model can be formulated as 
\vspace{-2mm}
$$minimize~objective~~K (Eq.~\text{5})$$
\vspace{-6mm}
$$s.t.~ constraint~ Eqs.~(\text{2, 3, 4})$$

\subsection{Optimizing Circuit Depth for Subgraph Generation}
\textbf{Optimizing Targets.} In this section, we introduce our compilation method for the subgraphs. 
Unlike previous works that focus on optimizing either the minimal emitter resource~\cite{li2022photonic} or \#CNOT~\cite{lin2024graphiq},
we propose a novel compilation method emphasizing the suppression of photon loss, informed by the practical constraints of current hardware platforms.
Our optimization targets are outlined below:

\paragraph{Number of emitter-emitter CNOTs}
Considering gates durations on hardware, emitter-emitter CNOTs are typically more challenging to implement than other operations.
As discussed in the quantum dot example in Section~II.B, the emitter-emitter CNOT time is more than $10\times$ longer than that of the emitter-photon CNOT,
constituting the primary contributor to the overall circuit duration.

\paragraph{Photon loss duration}
Another critical problem in graph state generation is the loss of photon qubits.
From the moment a photon is emitted until the entire graph state is fully generated, the photon accumulates loss, reducing the fidelity of the final graph state.
Minimizing the existing time of photons can suppress the probability of photon loss.
Specifically, given two circuits with the same depth (duration), we prefer the circuit that delays photon emission operations as much as possible.

\vspace{0.5mm}
\textbf{Compilation Strategy.} 
Based on the time-reversed model, our compilation of each subgraph involves the following steps: 

~\noindent\circled{1} 
We search for the graph state reduction sequence, utilizing the reduction operations detailed in Section~II.C, considering the maximum number of available emitter $ne_{limit}$. A heuristic-based depth-first searching is performed to find reduction sequences requiring the minimal number of \textit{emitter disentangle} operations, which corresponds to circuits with minimal emitter-emitter CNOTs. The degree of vertices is used as heuristic, prioritizing the reduction of lower-degree vertices to ensure that more entanglements are created by photon emissions rather than emitter-emitter CNOTs. Additionally, we truncate the operation sequences with a high number of emitter-emitter CNOTs to reduce the searching overhead.
    
~\noindent\circled{2} 
After obtaining a set of candidate sequences with minimal number of emitter-emitter CNOTs, these sequences are translated into circuits, and the one with the minimal photon loss duration is selected. Specifically, the average photon loss duration ($\overline{T}_{loss}$) is calculated by:
$$
    \overline{T}_{loss}=\frac{1}{n}\sum_{p=p_1,...,p_{n}}M_{circ\_end}-M_{emit}(p)
$$
where $M_{\text{circ\_end}}$ represents the circuit ending time, and $M_{\text{emit}}(p)$ represents the emission time of photon $p$.

\vspace{0.5mm}
\textbf{Flexible Resource Constraint.} We adopt a flexible resource constraint, enabling the generation of optimal circuits under varying numbers of available emitters.
For a subgraph $g$, the minimal number of required emitters, $ne_{\text{min}}$, is calculated using entanglement entropy theory~\cite{li2022photonic}. 
The subgraph is then compiled with different available emitter limits: $ne_{\text{limit}} = ne_{\text{min}}, ne_{\text{min}} + 1, ne_{\text{min}} + 2$, respectively, to generate circuits for each scenario.
This approach enhances parallelism within the subgraph when additional emitter qubits are available, facilitating qubit reuse, which will be discussed further in the next subsection.

\subsection{Subgraph Recombination \& Circuit Scheduling}
\begin{figure}[th]
% \vspace{-5mm}
    \centering
    \includegraphics[page=1,width=.48\textwidth]{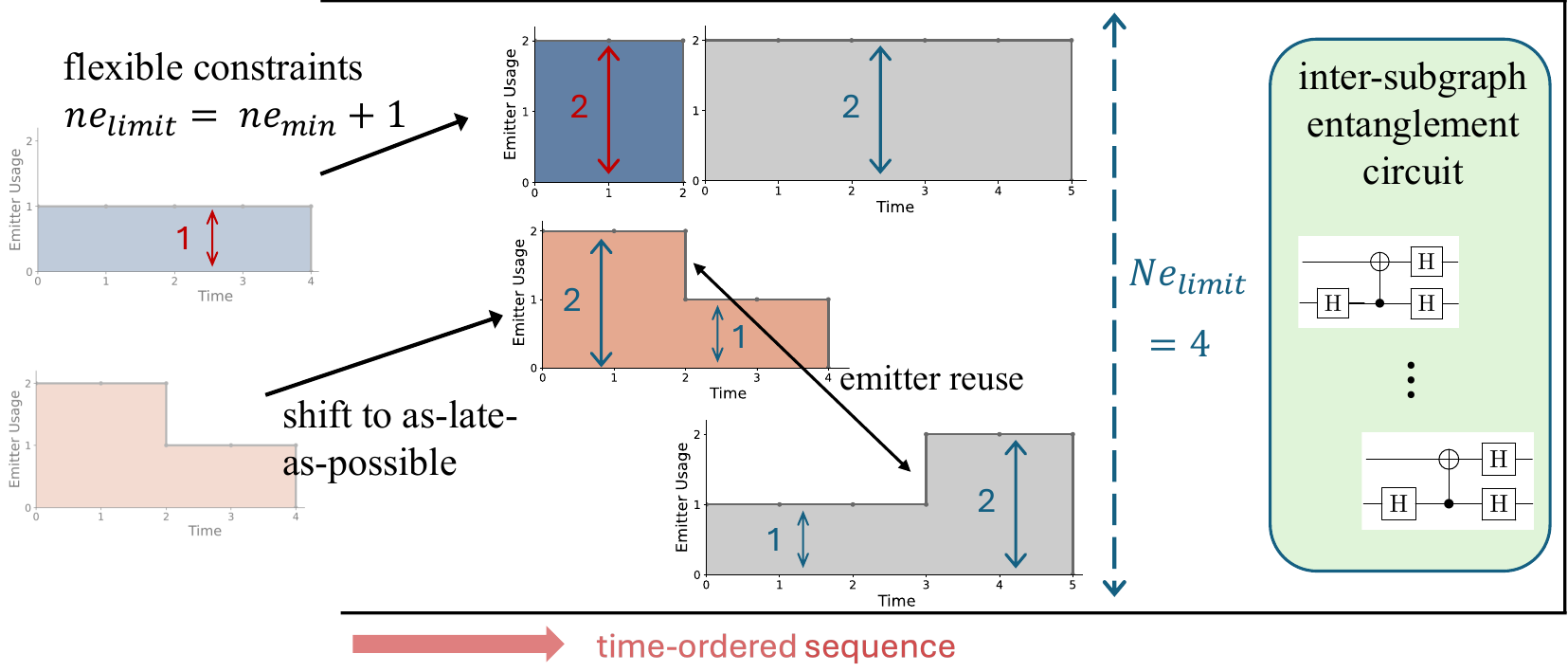}
    \caption{Our scheduling scheme. The blocks are emitter usage curve for each subgraph, and the width is restricted by total emitters $Ne_{limit}$.} 
    \label{fig:scheduling}
\vspace{-3mm}
\end{figure}

Finally, we combine the generation circuits corresponding to each subgraph and schedule these subcircuits. In this step, inter-subgraph entanglements are added to the overall circuit to connect all components. This process is illustrated in Figure~\ref{fig:scheduling}.

\textbf{Circuit Scheduling Strategy.}
Our scheduling scheme adopts the \textit{as-late-as-possible} strategy from Qiskit~\cite{qiskit2024}, appending subgraph circuits starting from the latest possible time and working gradually backward.
When selecting circuits to schedule, we prioritize subcircuits with more photon qubits and shorter execution times (placing them later),
and deprioritize subcircuits with fewer photon qubits and longer execution times (placing them earlier).
The priority coefficient for a subgraph circuit is calculated as: 
$P_{c}=\frac{n^c_p}{T_c}$, 
where $n^c_p$ is the number of photons in it, and $T_c$ is its duration.
This strategy ensures that more photons are emitted as late as possible, minimizing the time lasting for photon loss.

\textbf{Emitter Qubit Reuse.} The timestamp for a subgraph circuit $c$ is determined based on its emitter qubit usage curve. 
When inserting $c$ into the timeline of overall circuit $C$, we shift $c$ as late as possible, while ensuring that the total emitter usage in $C$ does not exceed the maximum emitter limit $Ne_{\text{limit}}$ at any moment. 
This approach is analogous to playing Tetris, where the pieces (subcircuits $c$) are shifted downward (later in time) as far as possible,
while being constrained by the walls (the $Ne_{\text{limit}}$), as shown in Figure~\ref{fig:scheduling}.

\textbf{Full Utilization of Emitter Qubits.}
To maximize the circuit parallelism in $Ne_{limit}$ constraint, we aim to utilize any remaining (unused) emitter qubits at each time slot.
This approach works in conjunction with the \textit{flexible resource constraint} policy in Section~IV.B.
For example, if no additional subgraph circuit $c$ can fit within a given time period, but the total emitter usage does not fully reach $Ne_{\text{limit}}$,
we relax the $ne_{\text{limit}}$ constraint from $ne_{\text{min}}$ to $ne_{\text{min}} + 1$ for the circuits already scheduled in that period.
This ensures full utilization of emitter qubit resources, ultimately reducing circuit duration.

\begin{figure}[b!]
\vspace{-5mm}
    \centering
    \includegraphics[page=1,width=.43\textwidth]{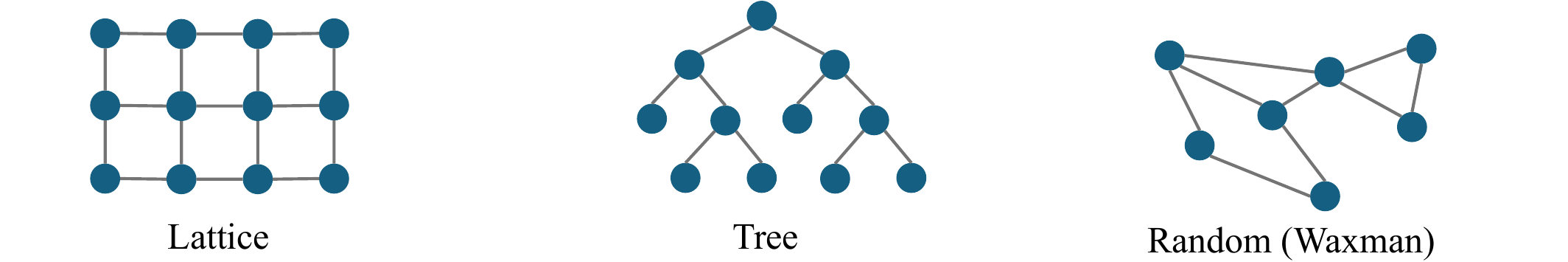}
    \caption{The types of graph in our benchmarks.} 
    \label{fig:graphtypes}
% \vspace{-3mm}
\end{figure}

\begin{figure*}[t]
    \centering
    \begin{subfigure}[b]{0.66\columnwidth}
        \centering
        \includegraphics[width=1\columnwidth]{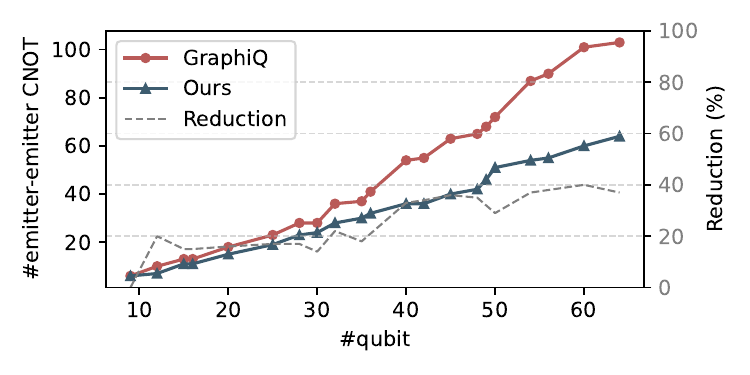}
        \vspace{-6mm}
        \caption{\#CNOT - lattice graph}
        % \label{fig:s_cut}
    \end{subfigure}
    \hfill
    \begin{subfigure}[b]{0.66\columnwidth}
        \centering
        \includegraphics[width=1\columnwidth]{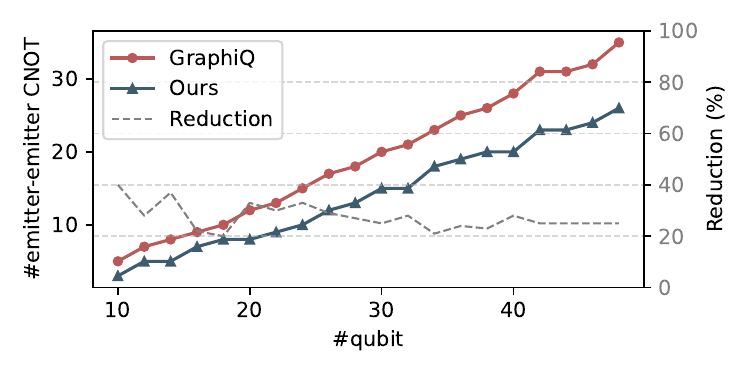}
        \vspace{-6mm}
        \caption{\#CNOT - tree graph}
        % \label{fig:s_depth}
    \end{subfigure}
    \hfill
    \begin{subfigure}[b]{0.66\columnwidth}
        \centering
        \includegraphics[width=1\columnwidth]{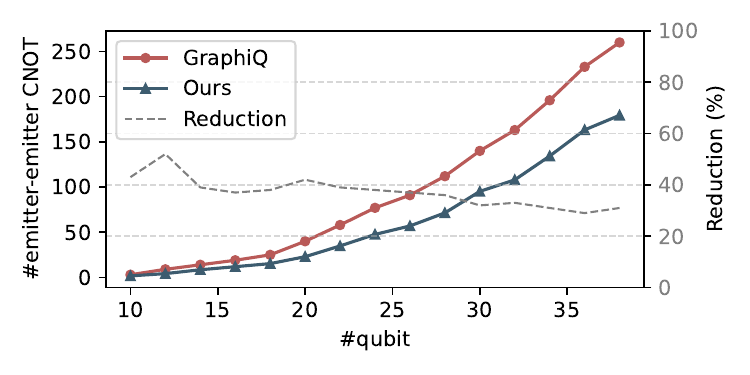}
        \vspace{-6mm}
        \caption{\#CNOT - random graph}
        % \label{fig:s_fidelity}
    \end{subfigure}
    \vspace{3mm}
    \begin{subfigure}[b]{0.66\columnwidth}
        \centering
        \includegraphics[width=1\columnwidth]{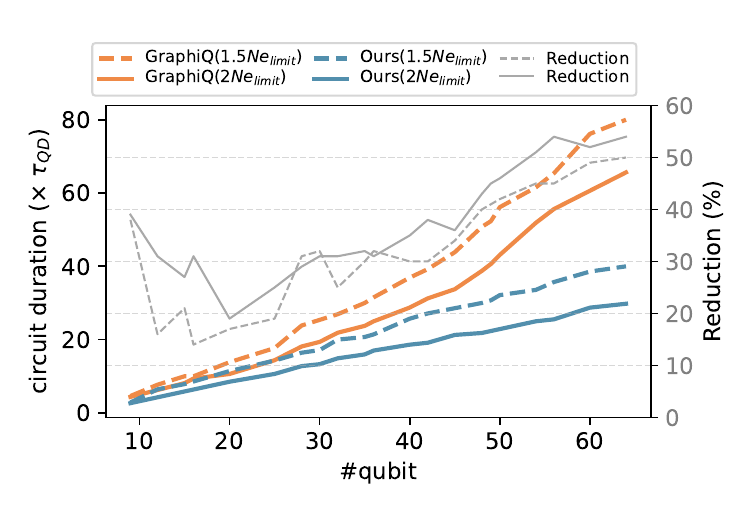}
        \vspace{-6mm}
        \caption{circuit duration - lattice graph}
        % \label{fig:s_cut}
    \end{subfigure}
    \hfill
    \begin{subfigure}[b]{0.66\columnwidth}
        \centering
        \includegraphics[width=1\columnwidth]{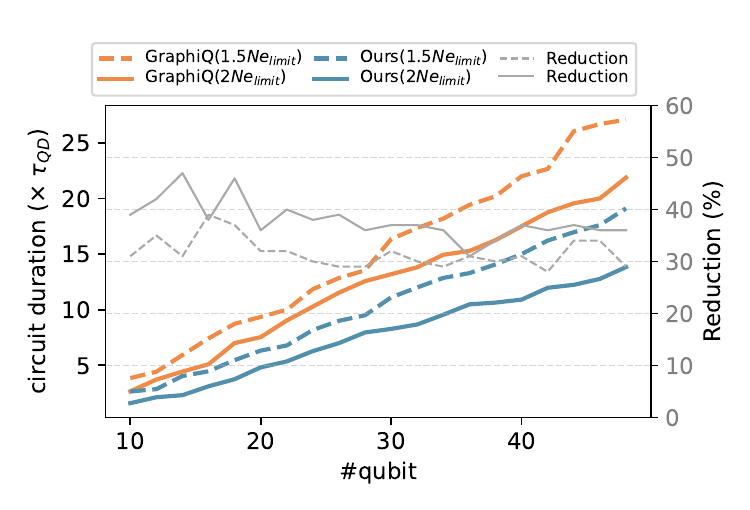}
        \vspace{-6mm}
        \caption{circuit duration - tree graph}
        % \label{fig:s_depth}
    \end{subfigure}
    \hfill
    \begin{subfigure}[b]{0.66\columnwidth}
        \centering
        \includegraphics[width=1\columnwidth]{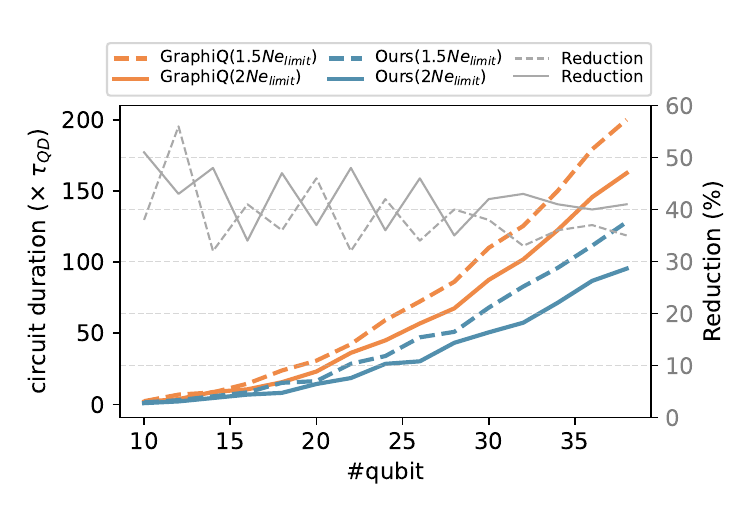}
        \vspace{-6mm}
        \caption{circuit duration - random graph}
        % \label{fig:s_fidelity}
    \end{subfigure}
    \vspace{-3mm}
    \caption{Evaluation: reduction of emitter-emitter CNOTs and circuit duration.}
    \vspace{-5mm}
    \label{fig:eval}
\end{figure*}

\section{Evaluation}
We evaluate our framework on various benchmarks using several hardware-realistic metrics and compare our results with the state-of-the-art graph state generation framework, GraphiQ~\cite{lin2024graphiq}.
\subsection{Experimental Setup}
\textbf{Benchmarks. }
The graph states we tested (Figure~\ref{fig:graphtypes}) are selected from %frequently used
popular quantum applications: 
1) \textit{Lattice graph} is a two-dimensional (2D) square grid structure, which is basic element for MBQC~\cite{li2023minimizing}; 
2) \textit{Tree graph} is a connected acyclic graph, which is particularly useful as it constitutes the quantum routers in quantum random access memory (QRAM)~\cite{xu2023systems}, while it's also vital to the tree code~\cite{azuma2015all} in quantum error correction;
3) \textit{Random graph} is generated by the Waxman graph structure~\cite{waxman1988routing}, which covers most of the possible communication topologies for distributed quantum computing and quantum network~\cite{huang2024space}.

\textbf{Experiment Platform. }
All compilation and simulation are executed on a CentOS 7 server featuring dual Intel Xeon 8255C CPUs (48 logical cores) and 384GB memory. The Python version is 3.10.0. 
Our framework is built based on GraphiQ\cite{lin2024graphiq} and Qiskit~\cite{qiskit2024}.

\textbf{Hardware Model.}
We select the silicon quantum dot (QD) emitter~\cite{russo2018photonic} as our hardware model for simulation. However, it can be easily adapted to other hardware platform, e.g. SiV color centers~\cite{stas2022robust}, NV color centers~\cite{choi2019percolation} and Rydberg atoms~\cite{yang2022sequential}, just by changing the configurations of gate characteristic.
In QDs hardware, an emitter-emitter CNOT is realized by electron exchange coupling between two QDs, with coefficient $J$ as the exchange interaction strength. Within a time period $\tau_1=\pi/2J$, the interaction Hamiltonian yields a $\sqrt{SWAP}$ gate, enabling the entangling between emitters. 
Then, a emitter-emitter CNOT consists of two $\sqrt{SWAP}$ interleaved by two $R_x \otimes R_x$, while each of them take $\tau_1$ as evolution time. 
Hence, we represent the period of emitter-emitter CNOT as $\tau_{QD} = 4\tau_1 = 2\pi/J$. 
The timescale of $\tau_{QD}$ depends on $J$ in hardware system, while it is commonly set to $2\pi \times 1$GHz and thus $\tau_{QD}=1$ns. 
On the other hand, we set the time period for photon emission as $0.1\tau_{QD}$ based on cavity enhancement~\cite{kelaita2016hybrid}.

\textbf{Algorithm Configuration \& Baseline.}
In our framework, we use Gurobi~\cite{gurobi} as the MIP solver for graph state partitioning. For the graph partitioning search algorithm in Section~IV.A, we set a 20-minute timeout for MIP solver, using the best solution found within this time. In MIP model, we limit the size of subgraph $g_{max}=7$ and maximum length of LC sequence to $l=15$. Note that the compilation runtime for each subgraph is relatively low ($<100$ seconds), owing to the constrained subgraph size. For baseline GraphiQ, we select its \textit{AlternateTargetSolver} and modify it to include a 30-minute timeout, instead of searching exhaustively. 

\begin{figure}[b]
\vspace{-4mm}
    \centering
    \includegraphics[page=1,width=.48\textwidth]{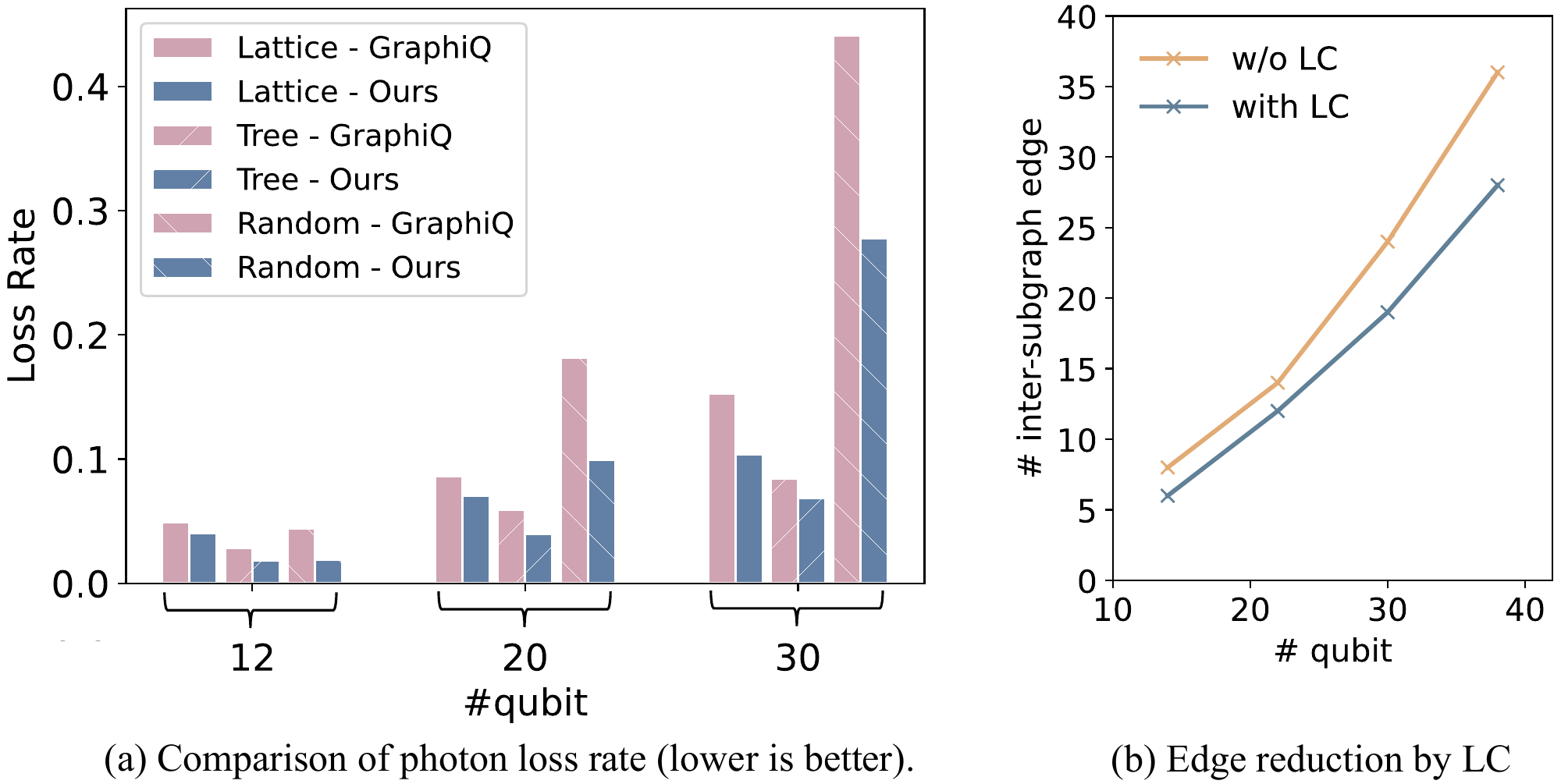}
    \caption{Evaluation: photon loss suppression and LC optimization.} 
    \label{fig:photonloss}
\vspace{-5mm}
\end{figure}

\subsection{Result}
\subsubsection{\# Emitter-emitter CNOTs}
As the emitter-emitter CNOT being a critical component of graph state generation circuit, we compare its count in the compilation of graph state in varying sizes. 
The results are shown in Figure~\ref{fig:eval}.(a)-(c) for three types of graphs.
Our compilation framework consistently reduces the \#CNOT compared to GraphiQ across all graph states, with average reductions of 25\%, 28\%, 37\% (maximum 40\%, 39\%, 52\%) for lattice, tree, and random graphs, respectively.
Furthermore, as the size of the graph state increases, the reduction rate grows for lattice graphs and remains steady for the others, demonstrating the scalability of our method.

\subsubsection{Circuit Duration}
The time unit we use for calculating duration is $\tau_{QD}$ introduced in last subsection, with its realistic value dependent on hardware.
We compare circuit durations under two emitter resource settings: $Ne_{limit} = 1.5Ne_{min}$ and $Ne_{limit} = 2Ne_{min}$ ($Ne_{min}$ is the minimal \#emitter required for target graph). 

Our results in Figure~\ref{fig:eval}.(d)-(f) show that our compilation framework reduces circuit duration significantly.
For $Ne_{\text{limit}} = 1.5Ne_{\text{min}}$, the average reduction percentages are 33\%, 32\%, and 39\% (maximum 50\%, 39\%, and 56\%).
For $Ne_{\text{limit}} = 2Ne_{\text{min}}$, the average reductions are 38\%, 38\%, and 43\% (maximum 54\%, 47\%, and 51\%).

For the same graph state, circuit duration reduction rates are consistently higher than \#CNOT reduction rates.
Additionally, the duration reduction rate generally increases when the emitter resource setting $Ne_{\text{limit}}$ is raised from $1.5Ne_{\text{min}}$ to $2Ne_{\text{min}}$.
This highlights the benefits of circuit parallelism, with its impact becoming more pronounced as more emitter resources are made available.

\subsubsection{Photon Loss}
We assess the photon loss rate of final graph state generated by circuit, based on realistic hardware settings. Given the electron spin coherence time $T_2\sim 1$s, we set the photon loss rate to 0.5\% per time period $\tau_{QD}$. Our simulations ($Ne_{limit} = 1.5Ne_{min}$) are performed using GraphiQ, and the results are given in Figure~\ref{fig:photonloss}.(a). Our framework improves the baseline by $\times$1.3, $\times$1.4, and $\times$1.9 on average for three types of graph, respectively.

\vspace{0.2mm}
\subsubsection{Local Complementation}
We compare the average number of inter-subgraph edges with and without local complementation optimizing in Figure.\ref{fig:photonloss}.(b) (by setting LC sequence $l=15$ or $0$), to illustrate the benefits of local complementation. 
In this experiment, we use random (Waxman) graphs as the benchmark.

\section{Conclusion}
We propose a novel compilation framework for emitter-photonic graph state generation, leveraging a divide-and-conquer strategy.
The framework partitions the graph state into subgraphs, compiles them separately, and then recombines them using circuit scheduling. 
Tested on various graph states, our framework shows reductions in \#CNOT and circuit duration, taking hardware constraints into account.

% \section*{Acknowledgment}
% We thanks Liang Jiang for helpful discussion and we acknowledge support from the ARO MURI (W911NF-21-1-0325) and NSF (ERC-1941583). Also, we thanks the reviewers of DAC25 for their invaluable feedback.

\bibliographystyle{IEEEtran}
\bibliography{references}

\end{document}